# Usage and Impact of ICT in Education Sector; A Study of Pakistan

M. Wasif Nisar, Ehsan Ullah Munir and Shafqat Ali shad

Department of Computer Science Comsats Institute of Information and Technology
Wah Cantt, Pakistan.

**Abstract:** In many countries, information and communication technology (ICT) has a lucid impact on the development of educational curriculum. This is the era of Information Communication Technology, so to perk up educational planning it is indispensable to implement the ICT in Education sector. Student can perform well throughout the usage of ICT. ICT helps the students to augment their knowledge skills as well as to improve their learning skills. To know with reference to the usage and Impact of ICT in Education sector of Pakistan, we accumulate data from 429 respondents from 5 colleges and universities, we use convenient sampling to accumulate the data from district Rawalpindi of Pakistan. The consequences show that Availability and Usage of ICT improves the knowledge and learning skills of students. This indicates that existence of ICT is improving the educational efficiency as well as obliging for making policies regarding education sector.

## INTRODUCTION

The awareness of primary school children's is the linkage between ICT and the way they learn within the situation of a school that has been particularly successful in integrating ICT into the curriculum. Incorporating teachers into the research design has the advantage of aiding the recall of particular learning episodes which, with children of this age, are a fertile ground for insights into the teaching and learning dynamic (Goodison, 2002). The teachers mainly focus on the development of technical ICT skills, whereas the ICT curriculum centers on the integrated use of ICT within the learning and teaching process. The potential value of a school-based ICT curriculum that 'translates' the national ICT-related curriculum into an ICT plan as part of the overall school policy (Tondeur, Braak, Valcke., 2006). Despite huge efforts to position information and communication technology (ICT) as a central tenet of university teaching and learning, the fact remains that many university students and faculty make only limited formal academic use of computer technology (Selwyn, 2006).

Intersections picturing the new technique as partly changing the situation for teaching, learning and collaboration between colleagues. Changed roles because of ICT competence raise questions about the importance of systematic ICT features within teacher education. Many of the newly qualified teachers wish they had more knowledge about ICT and related techniques (Andersson, 2006). The relationship between changes in ICT investment and changes in educational performance in Local Education Authorities (LEAs). In contrast with most previous studies in the economic literature, discover the evidence for a positive impact of ICT investment on educational performance in primary schools. This provides an interesting parallel to the existing work that does not find beneficial effects for pupils and to the related work on firms where there is evidence that ICT investment enhances firm productivity (Machin, McNally, Silva., 2007).

PowerPoint and other visual technologies have become persistent in schools. Adoption of these technologies is perceived as a necessary - or, at the very least, an educationally appropriate, even though systematic examination of their use is relatively recent (Reedy, 2008). Teacher use of computer mediated communication (CMC) in future will only be a success if the link between the possibilities of CMC and teacher practice becomes clearer. Language teachers seem to start using CMC, even though no evidence was found that these teachers have a more positive degree of technological innovativeness. The challenge for teachers is to integrate CMC not only for supportive tasks, but for teaching purposes as well. Nevertheless, computer networks offer enormous challenges for teaching and learning (Braak, 2000).

In powerful learning environments, rich contexts and authentic tasks are presented to pupils. Many teachers apply several elements of powerful learning environments in their classes. Chances of using open-ended ICT applications, which are expected to contribute to the power of learning environments, were greater with teachers who created powerful learning environments for their pupils, and when there were more computers available to pupils. Moreover, teachers' skills with regard to the use of ICT as a means to support powerful learning environments should be fostered (Smeets, 2004). The Jordanian Ministry of Education has given priority to the social and vocational rationales in launching the computer literacy and awareness course. Throughout central and direct supervision on schools, the Directorate is able to look after issues related to hardware and software, maintenance and teacher training (Tawalbeh, 2001).

**Corresponding Author:** M. Wasif Nisar, Department of Computer Science Comsats Institute of Information & Technology
Wah Cantt, Pakistan.
E-mail: wasifnisar@gmail.com

The issue is addressed through interpretive in a UK secondary school where almost all staff is now using ICT to enhance and extend learning in their subject areas (Tearle, 2003). Pupils' engagements with ICTs to be often perfunctory and unspectacular, especially within the school setting, where the influence of year group and school attended are prominent. There was a strong sense of educational uses of ICTs being constrained by the nature of the schools within which 'educational' use was largely framed and often situated (Selwyn, Potter, Cranmer; 2008). Children would now appear to have demands and expectations beyond the tokenistic 'go on the computer' as an end in itself. Thus teachers should strive for constructing meaningful and really useful opportunities for children to use computers and, therefore, stimulating continued desire to use ICT in school (Selwyn, Bullon., 2000).

The combined impact of both teacher and school characteristics was explored through a multilevel analysis. Besides the importance of school characteristics, the results reveal differential effects of specific characteristics on specific types of computer use. Variables at teacher and at school level are related to different types of computer use (Tondeur, Valcke, Braak., 2008). In order to enable all pupils to realize their potential, facilities and strategies have to be reviewed and developed to ensure that their learning experience is appropriate. The generic capability of ICT can enhance and enable the learning of pupils with special educational needs (SEN), but do not purport to consider specifically the specialized technology required for certain manifestations of special need. ICT certainly has much potential to enhance the teaching of all four skills in language learning, and to make skills available to a wider range of pupils (Meiring, Norman., 2005). Multilevel evaluation of professional development was shown to be robust for ICT teacher training, including a significant correlation between the views of experts and those of teachers. A second paper delves deeper to describe and contrast the highest and lowest-rated approaches to ICT teacher training (Davis, Preston, Sahin., 2008).

Development in the use of information and communications technology (ICT) in education of children with a visual impairment. It is argued that information and communications technology has a valuable role in providing opportunities for children with a visual impairment to participate more fully in education (Douglas, 2001). The ICT coordinators were both knowledgeable and enthusiastic about ICT use and this placed them in a good leadership position to guide and implement ICT use in the school. A full-time ICT coordinator is essential if ICT is to be successfully integrated into the school curriculum (Lai, Pratt., 2004).

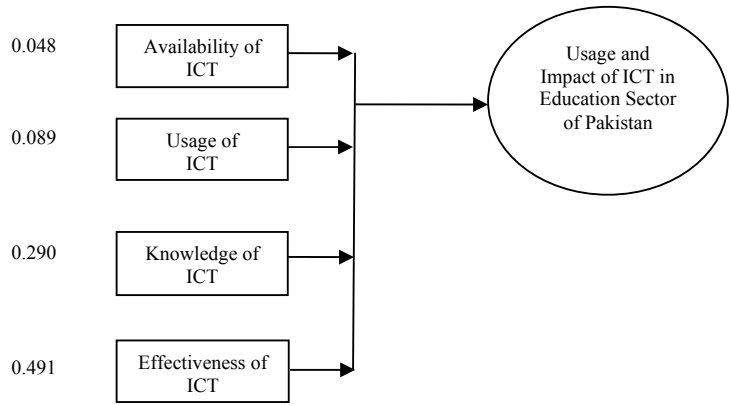

Theoretical Framework.

***Econometric Equation:***

We are going to study that how the Usage & Impact of ICT in Education sector of Pakistan is affected by availability, usage, knowledge and effectiveness of ICT. The other factors are going to be constant. The hypothesis is that these factors effect on Usage & Impact of ICT in education sector.

Econometric equation is as follow:

Usage and Impact of ICT = $\alpha$ + $\beta_1$ (Availability of ICT) + $\beta_2$ (Usage of ICT) + $\beta_3$ (Knowledge of ICT) + $\beta_4$ (Effectiveness of ICT) + $\mu$

Hypothesis:

$H_0$: Availability of ICT, usage of ICT, knowledge of ICT, Effectiveness of ICT does affect the relationship of Usage and impact of ICT in Education sector.

$H_1$: Availability of ICT, usage of ICT, knowledge of ICT, Effectiveness of ICT does affect the relationship of Usage and impact of ICT in Education sector.

***Methodology:***

To check the usage and impact of ICT in education sector we exploit four independent variable i.e. Availability of ICT, usage of ICT, knowledge of ICT, and Effectiveness of ICT.

*Availability of ICT:*
    The first variable that we exploit to check the usage and impact of ICT in education sector is Availability of ICT. To ensure how this variable help us to find out that what sort of Availability of ICT in education sectors of Pakistan. Respondent's reactions were evaluated at five likert scale and questions were:
(1) Availability of Well-equipped IT lab in university/college.
(2) Every time you use the Computer lab, internet is available for you.
(3) Availability of Multimedia during lectures.
(4) Availability of Digital Library in Computer Labs.

*Usage of ICT:*
    The second variable that we exploit to check the usage and impact of ICT in education sectors of Pakistan is usage of ICT. To check how this variable help us to find out that how can students use the latest technology in their studies. Respondents' reactions were evaluated at five likert scale and questions were:
(1) Usage of latest technology of ICT in university/college.
(2) Use of Multimedia device rather than White/Black board.
(3) Usage of Internet for doing assignments and projects rather than books or library.
(4) Usage of wireless communication in university/college.

*Knowledge of ICT:*
    The third variable that we exploit to check the usage and impact of ICT in education sectors of Pakistan is Knowledge of ICT. To check how this variable help us to find out the knowledge of the students regarding the use of ICT and how this knowledge helpful for the student at educational level. Respondent's reactions were evaluated at five likert scale and questions were:
(1) IT in Education provides information to operate different devices.
(2) IT in education sector provides knowledge that would be helpful at the professional level.
(3) IT helps to produce the productive knowledge to students related to their studies

*Effectiveness of ICT:*
    The last and fourth independent variable that we exploit to check the usage and impact of ICT in education sectors of Pakistan is Effectiveness of ICT. To check how this variable help us to find out the Effectiveness of ICT in education sector of Pakistan. Respondent's reactions were evaluated at five likert scale and questions were:
(1) Due to IT, Students can enhance their learning skills.
(2) IT provides vast knowledge to students through Internet.
(3) Use of Digital Projectors helps the students for better learning.
(4) IT can be used to enhance educational efficiency at the local, regional and national level.

*Usage and Impact of ICT in Education sector:*
    This dependent variable enlighten that how the student can use the latest technology of Information communication technology(ICT) in their studies and what are its impact in education sector of Pakistan. Respondent's response were estimate on five likert scales and questions were:
(1) IT can be used to enhance educational planning.
(2) Due to IT students delivers better results.
(3) IT can improve the knowledge skills of students.
(4) IT brings positive effect in Education sector of Pakistan.
(5) IT is efficiently used in education sector of Pakistan

*Sampling:*
    We have conducted research to stumble on the Usage and Impact of ICT in Education sector of Pakistan. For this intention we have made different samples using convenient sampling; a type of non-probability sampling. We have used questionnaire as our data collection technique for our research. Interval scale is used for formulating the questions we have used five likert scales starting with strongly disagree to strongly agree in our questionnaire.

Gender

|        |     |
|--------|-----|
| Male   | 198 |
| Female | 231 |
| Total  | 429 |

Nominal scale is being used to check the total number of male and female.

Age

|  |  |
|---|---|
| 15-20 | 339 |
| 20-30 | 90 |
| 30-40 | 0 |
| 40-above | 0 |
| Total | 429 |

Nominal scale is being used to check the age of Students.

For our research we conducted the survey from District Rawalpindi using questionnaire technique for this we issued 500 questionnaires and we got back 429 questionnaires. So to check the response rate we perform following formula:

$$\frac{\text{Responded questionnaire}}{\text{Total questionnaires}} * 100$$

$$\frac{429}{500} * 100 = 85.8\%$$

Hence Response rate is 85.8%.

*Data Analysis and Interpretation:*

**Table1:** Descriptive Statistics.

| Variables | Mean | Std. Deviation | N |
|---|---|---|---|
| Usage and impact of ICT | 4.0280 | .83395 | 429 |
| Availability of ICT | 3.0000 | .99297 | 429 |
| Usage of ICT | 3.0699 | .98339 | 429 |
| Knowledge of ICT | 3.9441 | .79815 | 429 |
| Effectiveness of ICT | 3.8788 | .85884 | 429 |

This table shows that overall response of our sampling (mean) from different colleges and universities of Pakistan consist of 429 students. After collecting the data from these students, we analyze that usage and impact of ICT as a dependent variable is 4.0280, and regarding to this variable respondent deviate from their mean equal to 0.83395, and results lie from 3.19405-4.86195, which shows that response of students lie between uncertain to strongly agree. They are agree that usage of ICT brings a positive impact on the education sector of Pakistan. They are agree that the usage of ICT improve their knowledge skills and helpful to deliver better results. They are also agree that ICT is efficiently used in education sector of Pakistan and enhance their planning regarding education.

Availability of ICT as an independent variable is 3.00 and regarding to this variable respondents deviate from their mean equal to 0.99297, and result lie from 2.00703-3.99297, which shows that response of students lie between disagree to almost agree. This result shows that students are uncertain about the availability of well equipped IT lab, internet in computer lab, use of multimedia during lectures as well as digital library in computer lab.

Availability of ICT as an independent variable is 3.0699 and regarding to this variable respondents deviate from their mean equal to 0.98339, and result lie from 2.08651-4.05329, which shows that response of students lie between disagree to agree. This result shows that students are uncertain about the usage of latest technology and the usage of multimedia device during lectures. They are also uncertain about the usage of internet for doing assignments/projects and as well as the usage of wireless communication in their institutions

Knowledge of ICT as an independent variable is 3.9441 and regarding to this variable respondents deviate from their mean equal to 0.79815, and result lie from 3.14595-4.74225, which shows that response of students lie between uncertain to almost strongly agree. This result shows that students are almost agree that ICT provides information to operate different devices and help to produce the productive knowledge that related to their studies, they are also agree that ICT in education sector provides knowledge that helpful at the professional level.

Effectiveness of ICT as an independent variable is 3.8788 and regarding to this variable respondents deviate from their mean equal to 0.85884, and result lie from 3.01996-4.73764, which shows that response of students lie between uncertain to almost strongly agree. This result shows that students are almost agree that due to ICT students can enhance their learning skills, provide vast knowledge to them through internet and use of digital projector helps for better learning. It means that ICT can be used to enhance educational efficiency at local, regional and national level.

**Table 2:** Correlation.

| Variables | Usage and impact | Availability | Usage | knowledge | Effectiveness |
|---|---|---|---|---|---|
| Usage and impact of ICT | 1.000 | .048 | .089 | .290 | .491 |
| Availability of ICT | .048 | 1.000 | .543 | .136 | .082 |
| Usage of ICT | .089 | .543 | 1.000 | .231 | .154 |
| Knowledge of ICT | .290 | .136 | .231 | 1.000 | .348 |
| Effectiveness of ICT | .491 | .082 | .154 | .348 | 1.000 |

This table shows the interdependency of variables like how much dependent variable depends on independent variables. Table shows that one time change in availability of ICT brings 0.048 change in Usage & Impact of ICT, as a result we analyze that there is a positive relationship between availability of ICT and Usage & Impact of ICT in Education Sector. Table shows that one time change in Usage of ICT brings 0.089 change in Usage & Impact of ICT, as a result we analyze that there is a positive relationship between Usage of ICT and Usage & Impact of ICT in Education Sector. Table shows that one time change in Knowledge of ICT brings 0.290 change in Usage & Impact of ICT, as a result we analyze that there is a positive relationship between Knowledge of ICT and Usage & Impact of ICT in Education Sector. Table shows that one time change in Effectiveness of ICT brings 0.491 change in Usage & Impact of ICT, as a result we analyze that there is a positive relationship between Effectiveness of ICT and Usage & Impact of ICT in Education Sector.

**Table 3:** Model Summary.

| R | R Square | Adjusted R Square |
|---|---|---|
| .507 | .257 | .250 |

The model summary table shows that Multiple Correlation coefficient (R), using all the predictors simultaneously, is 0.507($R^2$=0.257) and the adjusted $R^2$ is 0.250, it shows that there is 25% of the variance in Usage & Impact of ICT in education sector can be predicted from Availability, Usage, Knowledge & Effectiveness of ICT.

**Table 4:** ANOVA.

| | Sum of Squares | df | Mean Square | F | Sig. |
|---|---|---|---|---|---|
| Regression | 76.570 | 4 | 19.143 | 36.710 | .000 |

The table shows that the value of F-test is 36.710(which is greater than 12), and the significance level is 0.000, So it indicates that this is a best fitted model for research of Usage & Impact of ICT in Education sector of Pakistan and this model is helpful for future research.

**Table 5:** Coefficients.

| Variable | Beta | t |
|---|---|---|
| Usage and Impact of ICT | | 8.233 |
| Availability | -.001 | -.029 |
| Usage | -.011 | -.212 |
| knowledge | .138 | 3.038 |
| Effectiveness | .444 | 9.924 |

This table shows the efficiency of Independent Variables with Dependent variable. The value of 't' for first independent variable i.e. Availability of ICT is -0.029 which shows that it is less efficient variable. Also the value of 't' for Usage of ICT is -0.212 which shows that it is also less efficient variable. So it means that if both Availability of ICT and Usage of ICT is increased then there is no change in Usage & Impact of ICT in Education sector of Pakistan.

Now the value of 't' for Knowledge of ICT is 3.038 which shows that it is more efficient variable. Also the value of 't' for Effectiveness of ICT is 9.924 which shows that it is most efficient variable. It means that if both the Knowledge and Effectiveness of ICT will increased then the dependent variable i.e. Usage & Impact of ICT in Education sector will also increased.

***Conclusion:***
The major finding of this research is that availability and usage of ICT is very essential to improve the educational efficiency of students. This indicates that availability of ICT in Education is supportive for the students to improve their learning skills as well as latest technologies of ICT are helpful for the students to better prepare their assignments and projects. Results also show that ICT can helpful to produce the productive knowledge of students related to their studies. Our findings suggest that more the availability and usage of ICT in education sector will increase then as a result more the efficiency of students will increase. Students were agree that ICT provides vast knowledge to students through internet and digital libraries, so it can helpful to enhance the educational efficiency at local, regional and national level. After analyzing all the results we conclude that ICT brings a positive impact on Education sector of Pakistan.